\documentclass[aps,prc,twocolumn,superscriptaddress,showpacs,amsmath,nofootinbib]{revtex4}
\usepackage{graphicx}
\usepackage{bm}

\begin{document}

\title{Search for particle-bound $^{26}$O and $^{28}$F in $p$-stripping 
reactions}

\author{A.~Schiller}
\email{schiller@nscl.msu.edu}
\affiliation{National Superconducting Cyclotron Laboratory, Michigan State 
University, East Lansing, MI 48824}
\author{T.~Baumann}
\affiliation{National Superconducting Cyclotron Laboratory, Michigan State 
University, East Lansing, MI 48824}
\author{J.~Dietrich}
\altaffiliation[Permanent address: ]{Institut f\"{u}r Kern- und Teilchenphysik,
Technische Universit\"{a}t Dresden, D-01069 Dresden}
\affiliation{National Superconducting Cyclotron Laboratory, Michigan State 
University, East Lansing, MI 48824}
\author{S.~Kaiser}
\altaffiliation[Permanent address: ]{Institut f\"{u}r Kern- und Teilchenphysik,
Technische Universit\"{a}t Dresden, D-01069 Dresden}
\affiliation{National Superconducting Cyclotron Laboratory, Michigan State 
University, East Lansing, MI 48824}
\author{W.~Peters}
\affiliation{National Superconducting Cyclotron Laboratory, Michigan State 
University, East Lansing, MI 48824}
\affiliation{Michigan State University, Department of Physics and Astronomy, 
East Lansing, MI 48824}
\author{M.~Thoennessen}
\affiliation{National Superconducting Cyclotron Laboratory, Michigan State 
University, East Lansing, MI 48824}
\affiliation{Michigan State University, Department of Physics and Astronomy, 
East Lansing, MI 48824}

\begin{abstract}
We have searched for particle-bound $^{26}$O and $^{28}$F isotopes in the 
reaction products of secondary $^{27}$F and $^{29}$Ne beams, respectively. No
events have been observed. Upper limits for the respective production cross 
sections by one-$p$-stripping reactions are established under the assumption 
that $^{26}$O and $^{28}$F are particle bound. Since the experimental upper 
limits are much lower than common estimates we conclude that neither $^{26}$O 
nor $^{28}$F are likely particle bound.
\end{abstract}


\maketitle

The existence of particle-bound isotopes with given numbers of protons and 
neutrons is one of the most fundamental questions in nuclear physics. 
Particularly intriguing is the fact that in projectile fragmentation reactions
$Z=8$ oxygen isotopes have been found only for neutron numbers up to $N=16$ 
\cite{FM96} while neighboring $Z=9$ fluorine isotopes can be found to up to at 
least $N=22$ \cite{SL99}. This means that one proton in the $sd$ shell can bind
as many as six more neutrons beyond the new magic number $N=16$ \cite{Br01}. 
Although modern theory with new effective interactions has been successful in
explaining this observation in terms of the effective single-particle energy of
the $\nu 0d_{3/2}$ orbital \cite{Br01} and how this energy is modified by the 
presence of one or more protons in the $\sigma\tau$-partner orbital 
$\pi 0d_{5/2}$ \cite{OF01,UO01}, one question remains. Since theory predicts 
$^{26}$O to be particle unbound by only about $\sim 20$~keV \cite{Br01,VZ05} 
and production cross sections in projectile fragmentation reactions decrease 
sharply with decreasing particle-separation energy of the desired reaction 
product \cite{LQ85,PG86}, there remains a possibility for $^{26}$O to be 
marginally bound and thus observable only in less violent types of reactions 
for which production cross sections do not depend strongly on 
particle-separation energies.

For this reason, we have investigated the $p$-stripping reaction of a 
radioactive $^{27}$F beam on a $\sim 146$-mg/cm$^2$-thick carbon target at beam
energies of $\sim 90$~MeV/u. The experiment was performed at the Coupled 
Cyclotron Facility of the National Superconducting Cyclotron Laboratory at 
Michigan State University. A $\sim 140$~MeV/u $^{48}$Ca beam was fragmented on 
an $\sim 850$-mg/cm$^2$-thick Be target. Reaction products from this target 
were selected using the A1900 fragment separator with a 
$\sim 1000$-mg/cm$^2$-thick acrylic wedge at the intermediate image. The 
momentum acceptance was limited to 1\%. Major impurities with roughly equal
intensities as the desired $^{27}$F beam were $^{29}$Ne and $^{30}$Na. The 
presence of $^{29}$Ne in the beam enabled us to also investigate the 
$p$-stripping reaction from this nucleus and to address the question whether 
$^{28}$F can be observed among the reaction products or not. The total number 
of incoming beam particles was $\sim 3\times 10^5$ and $\sim 5\times 10^5$ for
$^{27}$F and $^{29}$Ne, respectively. 

\begin{figure}[htb!]
\includegraphics[totalheight=5.6cm]{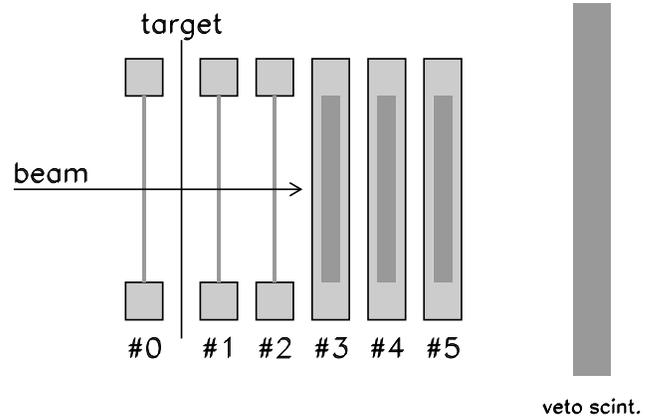}
\caption{Detector setup at the focal plane of the A1900 fragment separator for
isotopic identification of reaction products using the $\Delta E$--$E$ method. 
Diffraction-dissociation events where a light charged particle like a proton 
leaves the target can be discriminated against using the veto scintillation 
detector.}
\label{fig:stack}
\end{figure}

At the focus of the fragment separator, a stack of Si detectors was mounted 
which also contained the secondary carbon reaction target (see Fig.\ 
\ref{fig:stack}). The stack consisted of a $\sim 100$-$\mu$m-thick Si 
surface-barrier detector (\#0) to identify the incoming beam particles event by
event using the energy-loss time-of-flight technique, the secondary reaction 
target, two more $\sim 100$-$\mu$m-thick Si surface-barrier detectors (\#1,2), 
three $\sim 5000$-$\mu$m-thick Li-drifted Si detectors (\#3--5), and a 
scintillation veto detector. The oxygen and fluorine reaction products of 
interest are stopped in the final Si detector (\#5). Light charged particles 
like, e.g., protons from diffraction-dissociation events will make it into the
plastic scintillation detector. The setup is nearly identical to the setup in 
a previous experiment for measuring one-$p$-stripping cross sections from 
$^{24-26}$F \cite{TB03}. 

The Si detectors were calibrated using $^{19-22}$O and $^{21-25}$F isotopes of 
several known energies, beams of which were created in runs of the A1900 
fragment separator with different $B\rho$ settings and without the acrylic 
wedge. These beams were also used to calibrate the mass-indicator spectra which
are used to separate different isotopes created in the stripping reactions. In 
the data analysis, particles of interest were identified by their large energy 
deposition in the three thick detectors (\#3--5). Unreacted beam particles were
separated from lower-$Z$ reaction products by the somewhat smaller energy 
deposition of the latter in the first three detectors behind the target 
(\#1--3). The detailed pattern of energy deposition in the three thick 
detectors (\#3--5) was used to construct three different mass indicators of 
which the first two do not rely on any specific functional form of the range 
curve. Range curves, i.e., penetration depths as function of kinetic energy 
$R(E)$ are calculated for $^{26}$O and $^{28}$F in silicon and fitted by 
third-order polynomials. So-called thickness spectra can then serve as mass 
indicators. In our case, we used 
\begin{eqnarray}
\label{eq:massid1}
t_{3,4}&=&R(E_{\#3}+E_{\#4}+E_{\#5})-R(E_{\#5})\\
\label{eq:massid2}
t_4&=&R(E_{\#4}+E_{\#5})-R(E_{\#5})
\end{eqnarray}
which for the desired isotopes yield the correct thicknesses of the Si 
detectors \#3+\#4 and \#4 for $t_{3,4}$ and $t_4$, respectively. Lighter 
isotopes of the same element yield smaller 'effective' values for the 
thicknesses, since the range curve is not matched to them. By construction, 
thickness spectra give isotopic identification independent of kinetic energy. 
The third mass indicator relies on the parameterization $R(E)\propto E^\alpha$ 
of the range curve \cite{SI79} from which the following mass indicator is 
derived:
\begin{equation}
\ln(\alpha\,\Delta E)+(\alpha-1)\,\ln(E+c\,\Delta E)-\alpha\,\ln(300),
\label{eq:massid3}
\end{equation}
where $\alpha=a-b\,\Delta E/T$ and $T$ is the $\Delta E$-detector thickness in 
micrometer. With this mass indicator, we use Si detector \#4 and \#5 as 
$\Delta E$ and $E$ detector, respectively. With calibration of the parameters 
$a$, $b$, and $c$ within their proper ranges, all three mass indicators give 
very similar results.

\begin{figure}[htb!]
\includegraphics[totalheight=17.3cm]{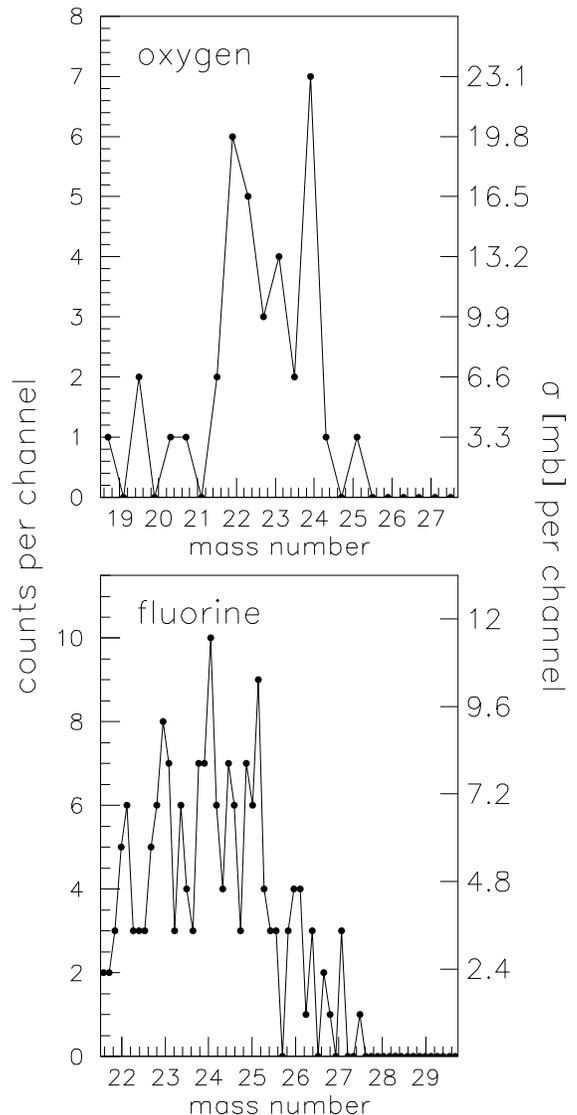}
\caption{Mass-indicator spectra for reaction products from proton stripping 
using a $^{27}$F (upper panel) and $^{29}$Ne (lower panel) beam.}
\label{fig:massid}
\end{figure}

Figure \ref{fig:massid} shows the mass-indicator spectra for reaction products
from proton stripping using a $^{27}$F and $^{29}$Ne beam, respectively. For
fluorine reaction products from the $^{29}$Ne beam, Eq.\ (\ref{eq:massid1}) 
gives the best result. For oxygen reaction products from the $^{27}$F beam, we
use Eq.\ (\ref{eq:massid3}). Moreover, for the latter reaction, we require (i) 
no hits in the veto scintillator which discriminates against 
diffraction-dissociation events where the stripped protons exit the target and
will deposit energy in the following Si detectors alongside the oxygen 
fragments, and (ii) an upper limit of $\sim 850$~MeV for energy deposition in 
detector \#4 in order to get rid of slow oxygen isotopes. The second condition 
is necessary since the calibration of the Si detectors and mass indicators is 
not well constrained for events where slow reaction products are almost stopped
in detector \#4, thus depositing most of their energy in there and very little 
in detector \#5\@.

No events corresponding to $^{26}$O or $^{28}$F are visible in the 
mass-indicator spectra. The presence of lighter isotopes can be explained by 
either sequential neutron decay after proton stripping events in which the 
residual nucleus is left in neutron-unbound states, or by more dissipative 
reactions like $1p\,xn$, $d$, or $t$ stripping. 

Given target thickness and the total number of incoming beam particles, each 
count in the fluorine mass-indicator spectrum equates a production cross 
section of $\sim 1.2$~mb, assuming $\sim 100$\% detection efficiency. Thus, the
non observation of $^{28}$F corresponds to an upper limit of its production 
cross section by proton stripping from $^{29}$Ne of $\sim 1.2$~mb. Production 
cross sections of lower-mass fluorine isotopes are in the order of 15--40~mb, 
in good agreement with earlier measurements in this mass region \cite{TB03}. 
The exact numbers should actually be reduced somewhat to account for reactions 
which occurred in the last layers of detector \#0 or the early layers of 
detector \#1\@. 

The production cross sections include events where the proton remains stuck in 
the target (knockout) and where it exits the target with roughly the same speed
as the fragment (diffraction dissociation). In the latter case, the protons 
will induce signals in the veto scintillator. For the production of fluorine 
isotopes, $\sim 40$\% of all events show coincident hits in the scintillator. 
Assuming that all of these events are related to diffraction-dissociation 
processes while all events without scintillator response are related to 
knockout processes, the total production cross section for each fluorine 
isotope will split up roughly as 40\% and 60\% between the two processes. 

In the case of oxygen isotopes, since we require no hits in the veto 
scintillator, diffraction-dissociation processes are excluded from the data. 
Assuming the same ratio between knockout and diffraction-dissociation events as
for the production of fluorine isotopes, each count in the mass-indicator 
spectrum equates a production cross section of $\sim 2.0$~mb for the knockout 
process alone, and of $\sim 3.3$~mb for knockout and diffraction-dissociation 
processes combined. The non observation of $^{26}$O corresponds therefore to an
upper limit of its production cross section by proton stripping from $^{27}$F 
of $\sim 3.3$~mb. We neglect here the effect of the cut on the maximum energy 
deposition in detector \#4, since events which would be affected by this cut 
would correspond to a quite dissipative production mechanism of $^{26}$O\@. 
However, this cut might impact the detection efficiency of lighter oxygen 
isotopes considerably. Production cross sections of lighter isotopes are in the
order of $\sim 30$~mb, where some uncertainty has to be attributed to the 
energy cut in the spectrum of Si detector \#4\@. Again, the production cross 
sections are very similar to the production cross sections of oxygen isotopes 
from proton stripping of lighter fluorine beams \cite{TB03}. However, 
production of oxygen isotopes due to proton stripping in the last layers of 
detector \#0 and the first layers of detector \#1 has been neglected. When 
taken into account, this effect reduces the quoted production cross sections 
somewhat. 

In conclusion, we have investigated the proton-strip\-ping reactions of 
$^{27}$F and $^{29}$Ne. No events of $^{26}$O or $^{28}$F have been observed 
which corresponds to upper limits of their production cross sections in the 
order of 3.3~mb and 1.2~mb, respectively. Production cross sections of lighter 
isotopes are in the order of 10--40~mb, comparable with earlier results from 
proton stripping of lighter fluorine beams. From the relatively low upper 
limits for production of $^{28}$F and $^{26}$O, we conclude that these isotopes
are not particle bound. In the case of $^{26}$O, this finding is at variance
with early theoretical estimates and in agreement only with more recent 
calculations using a modified shell-model interaction.

\end{document}